\documentclass[]{aa}
\usepackage{graphicx}
\usepackage[]{txfonts}
\usepackage{natbib}
\bibpunct{(}{)}{;}{a}{}{,} 

\newcommand{\MOST}{{\em MOST\/}}
\newcommand{\WIRE}{{\em WIRE\/}}
\newcommand{\GOLF}{{\em GOLF\/}}
\newcommand{\SOHO}{{\em SOHO\/}}
\newcommand{\VIRGO}{{\em VIRGO\/}}
\newcommand{\ms}{\mbox{m\,s$^{-1}$}}
\newcommand{\muHz}{\mbox{$\mu$Hz}}
\newcommand{\cms}{\mbox{cm\,s$^{-1}$}}
\newcommand{\acen}{\mbox{$\alpha$~Cen}}
\newcommand{\acena}{\mbox{$\alpha$~Cen~A}}
\newcommand{\acenb}{\mbox{$\alpha$~Cen~B}}

\begin{document}

\title{The non-detection of oscillations in Procyon by \MOST{}:\\ is it really
a surprise?}


\author{T.R. Bedding\inst{1} \and H. Kjeldsen\inst{2} \and 
F.~Bouchy\inst{3} \and
H.~Bruntt\inst{2,4} \and
R.P.~Butler\inst{5} \and
D.L.~Buzasi\inst{4} \and
J.~Christensen-Dalsgaard\inst{2} \and
S.~Frandsen\inst{2} \and
J.-C.~Lebrun\inst{6} \and
M.~Marti{\'c}\inst{6} \and
J.~Schou\inst{7}} 

\institute{
School of Physics, University of Sydney 2006, Australia
\and 
Department of Physics and Astronomy, University of Aarhus, DK-8000 Aarhus C,
	Denmark
\and
Laboratoire d'Astrophysique de Marseille, Traverse du Siphon BP8, 13376
Marseille Cedex 12, France
\and 
Department of Physics, US Air Force Academy, 2354 Fairchild Drive, CO
80840, USA 
\and 
Carnegie Institution of Washington, Department of Terrestrial Magnetism,
Washington, DC 20015-1305, USA
\and 
Service d'A{\'e}ronomie du CNRS, BP 3, 91371 Verri{\`e}res le Buisson,
France
\and 
W.W.~Hansen Experimental Physics Laboratory, Stanford University, 445 Via
Palou, Stanford, CA 94305-4085, USA
}

\titlerunning{Oscillations in Procyon}
\authorrunning{T.R.~Bedding et al.}


\date{}

\abstract{We argue that the non-detection of oscillations in Procyon by the
\MOST{} satellite reported by \citet{MKG2004} is fully consistent with
published ground-based velocity observations of this star.  We also examine
the claims that the \MOST{} observations represent the best photometric
precision so far reported in the literature by about an order of magnitude
and are the most sensitive data set for asteroseismology available for any
star other than the Sun.  These statements are not correct, with the most
notable exceptions being observations of oscillations in \acena{} that are
far superior.  We further disagree that the hump of excess power seen
repeatedly from velocity observations of Procyon can be explained as an
artefact caused by gaps in the data.  The \MOST{} observations failed to
reveal oscillations clearly because their noise level is too high, possibly
from scattered Earthlight in the instrument.  We did find an excess of
strong peaks in the \MOST{} amplitude spectrum that is inconsistent with a
simple noise source such as granulation, and may perhaps indicate
oscillations at roughly the expected level. 
}

\maketitle

\section{Introduction}

The \MOST{} satellite (Microvariability and Oscillations of Stars;
\citealt{WMK2003}) comprises a 15-cm telescope and CCD detector designed to
perform high-precision photometry on bright stars.  \citet[][hereafter
M04]{MKG2004} reported observations of the F5 star Procyon~A made almost
continuously over 32~days.  The Fourier amplitude spectrum showed no
evidence for oscillations and M04 concluded that if there are p-modes in
Procyon, they must have lifetimes less than 2--3 days and/or peak
amplitudes $<$15 parts per million.  We agree completely with this
conclusion but we do not agree with some of the other statements in that
paper, as outlined in the following sections.  For other discussion of
the issues raised in M04, see also \citet{E+T2004}.

\section{Lowest noise ever?}

We have measured the noise level in the \MOST{} amplitude spectrum from the
data plotted in Figure~3 of M04, which was provided to us in digital form
by the \MOST{} team.  We find that the noise is about 2\,ppm (parts per
million) at frequencies above 2\,mHz and 2.5 to 4.0\,ppm between 2.0 and
0.5\,mHz, in agreement with the analysis by the \MOST{} team (J. Matthews,
priv.\ comm.).

M04 described the \MOST{} result as ``the best photometric precision so far
reported in the literature \citep{GBK93,KvWR97,RBB2004} by about an order
of magnitude.''\footnote{Due to an error in typesetting, the reference
numbers in the text of \citet{MKG2004} do not match those in the list of
references.  The correct numbers are given in the Erratum, which we repeat
here for convenience: References 1 to 26 should be, respectively: 1,
10--16, 2, 3, 17--26 and 4--9.}  This statement is not correct.  The work
by \citet{GBK93} was a ground-based photometric campaign to monitor several
stars in the open cluster M67, most of which had spectral types similar to
Procyon.  Table~11 of that paper gives the final noise levels in the
amplitude spectra in the frequency range 2.5--4.5\,mHz (column 4).  Values
for the best ten stars (excluding Star~12, which is a red giant) are
6.4--9.7\,ppm.  These noise levels, obtained after one week, are certainly
higher than the 2\,ppm reported for the 32-day \MOST{} observations, but
only by factors of 3--5 and not an order of magnitude.  We also note
that the noise level for a given observing time is very similar for the two
data sets.

%

The observations by \citet{KvWR97} and \citet{RBB2004} cited by M04
targetted Ap stars and were indeed much less precise than the \MOST{} data.
However, the \citeauthor{RBB2004} observations were made from space using
the star tracker on the \WIRE{} satellite, and the same instrument achieved
far better precision (and good evidence for oscillations) on the solar-type
star $\alpha$~Cen~A \citep{S+B2001}.  Indeed, the noise level from those
observations was substantially below 1\,ppm and therefore considerably
better than that from the \MOST{} observations of Procyon.

M04 also stated that the duty cycle of 99\% for 32~days makes the
\MOST{} data the most sensitive data set for asteroseismology available
for any star other than the Sun.  While the duty cycle is certainly
tremendously good, one must also take the precision into account and such a
grand claim is certainly not justified.  The noise at high frequencies
(about 2\,ppm above 2\,mHz) is about half the peak solar signal.  Many
ground-based velocity campaigns over more than a decade have achieved
equivalent noise levels well below this on Procyon (e.g., 8\,\cms{} by
\citealt{MLA2004}, which is a third of the peak solar signal) and other
stars \citep[for recent reviews, see][]{B+C2003,B+K2003}.

Recent ground-based velocity observations of the \acen{} binary system are
particularly noteworthy.  Noise levels have reached 4.3\,\cms{} in \acena{}
\citep{B+C2002}, 3.8\,\cms{} in \acenb{} \citep{C+B2003} and, most
recently, 2.0\,\cms{} in \acena{} \citep{BBK2004}.  The latter
value is a mere tenth of the peak solar oscillation amplitude, and all
three campaigns produced data sets that are much more sensitive for
asteroseismology than the \MOST{} Procyon data.  This is reflected in the
fact that they actually succeeded in detecting oscillations in stars whose
amplitudes are substantially lower than Procyon's.  


\section{Previous work on Procyon}

Several radial-velocity studies of Procyon have detected an excess of power
around 0.5--1.5\,mHz \citep{BGN91,MSL99,BSB2002,KST2003,MLA2004,ECB2004}.
The reported peak amplitudes of 50--70\,\cms{} (not 1\,\ms, as quoted by
M04) are lower than predicted by theoretical models of oscillations
\citep{ChD+F83,HBChD99}.  Even before most of these observations were made,
\citet{K+B95} concluded that F-type stars, namely Procyon and several
members of M67, must oscillate with amplitudes less than has generally been
assumed.  This point has been raised several times since then
\citep{MSL99,S+H2000,B+K2003,MLA2004}.  The knowledge that Procyon
oscillates with an amplitude only 2--3 times solar, rather than
the theoretically predicted value of at least 4--5 times solar, has
therefore been in the literature for many years.  The \MOST{} upper limit
of 15\,ppm (four times solar -- see Sec.~\ref{sec.simulation}) is
consistent with this and is therefore hardly a surprise.

It should also be noted that the theoretical predictions are based on our
clearly incomplete understanding of stellar convection.  Indeed,
\citet{ChD+F83} noted that mixing-length theory overestimates the
convective flux in relatively hot stars and that predictions of oscillation
amplitudes may have to be reduced.  Interestingly, hydrodynamical
simulations of convection for a variety of stellar parameters by
\citet{SGT2004} also showed a substantial increase in the energy input from
convection to the modes with increasing effective temperature.  However,
\citeauthor{SGT2004} did not estimate the damping rate of the oscillations
and hence could not provide estimates of the resulting amplitudes.

Although several independent detections of excess power in Procyon have
been made from velocity measurements, it has been difficult to obtain a
power spectrum that shows a clear p-mode structure.  There is consensus
that the so-called large frequency separation is about 55\,\muHz{}
\citep{MMM98,MSL99,MLA2004,ECB2004}, which agrees well with theoretical models
\citep{BMM99,CDG99,DiM+ChD2001,PMB2002,KTM2004}, but there is not yet
agreement on the actual frequencies.  This is presumably due either to
insufficient high-quality data (i.e., confusion from aliases in the
spectral window) or to intrinsic properties of the star (i.e., short mode
lifetimes and/or modes shifted from the asymptotic relation by avoided
crossings).

\section{Could \MOST{} have detected oscillations?}
\label{sec.simulation}

For reference in the following discussion, we note that the strongest
oscillation modes in the Sun have velocity amplitudes of 20--25\,\cms{} and
intensity amplitudes in the blue-visual part of the spectrum of
4.5--5.0\,ppm.  Of course, these peak heights decrease when the observing
time becomes significantly longer than the typical lifetime of the
modes.\footnote{We follow the usual definition that the lifetime of a
damped oscillator is the time for the amplitude to decay by a factor
of~$e$.} Note that M04 quoted mode lifetimes for the Sun of days to weeks,
referring to \citet{T+F92}.  However, the reciprocal linewidths given in
Table~6 of \citet{T+F92} must be divided by $\pi$ to convert to mode
lifetimes, as discussed in section 4 of that paper.  The dominant modes in
the Sun have lifetimes of 2--4~days \citep[see also][]{CEI97}.

We must also keep in mind that intensity variations are a consequence of
temperature variations and, while the two are proportional, the constant of
proportionality depends on the effective temperature of the star
\citep[see][Eq.~5]{K+B95}.  In a star with the same physical oscillation
amplitude as the Sun (as measured in velocity) but with the effective
temperature of Procyon ($T_{\rm eff} = 6530$\,K), we would observe an
intensity amplitude at these wavelengths that is $T^2_{\odot}/T^2_{\rm eff}
= 0.78$ times the intensity amplitude we observe in the Sun.  Thus, the
upper limit of 15\,ppm in Procyon reported by M04 corresponds to a
``physical'' oscillation amplitude of 4 times solar, which translates to a
velocity amplitude of just below 1\,\ms.  A velocity amplitude of 60\,\cms,
as reported from ground-based Doppler measurements, corresponds to an
intensity amplitude of 9.5\,ppm, which is below the limit set by~M04.

\begin{figure*}
 \resizebox{\hsize}{!}{\includegraphics{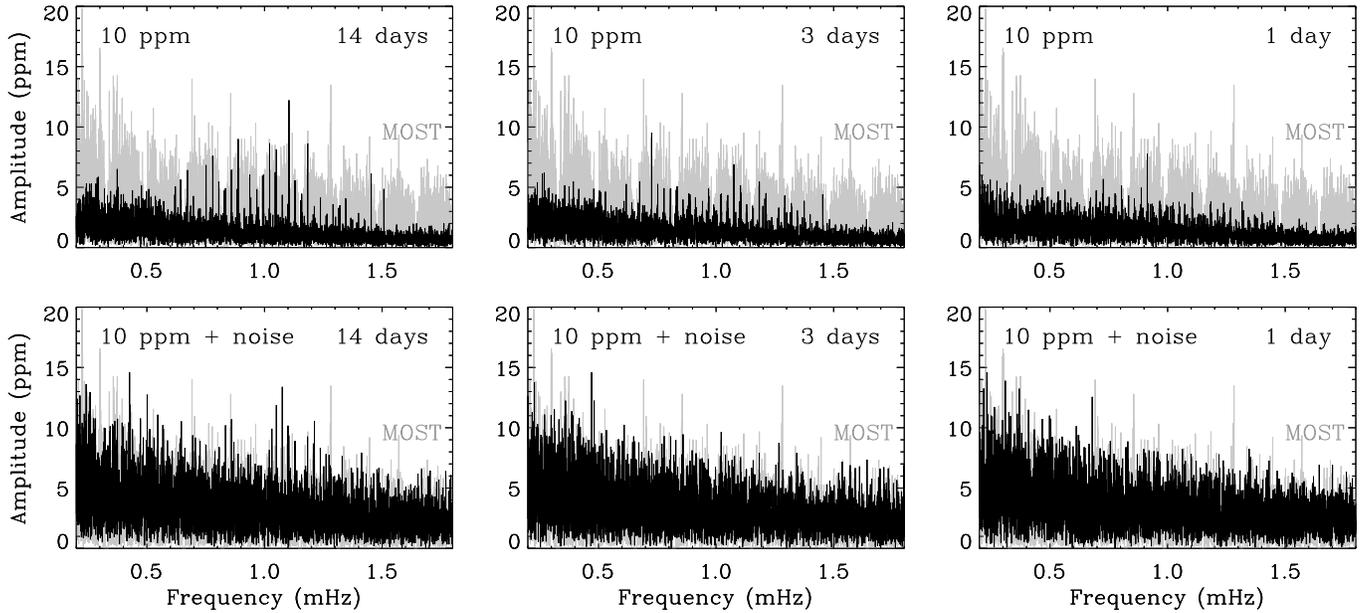}}
\caption{\label{fig.sim} The amplitude spectrum of Procyon observed by
\MOST{} (grey), compared with simulations (black) that included
oscillations and granulation at the expected levels.  The simulations were
made for three different mode lifetimes, both with and without instrumental
noise.}
\end{figure*}

To illustrate some of these points, Fig.~\ref{fig.sim} compares the \MOST{}
results with simulations.  In each panel, the amplitude spectrum in grey is
the spectrum published by M04 (the notches correspond to harmonics of the
satellite orbital frequency, where they filtered to removed power from
stray light artifacts).  The amplitude spectra shown in black are
simulations of the signal expected from Procyon if the p-mode oscillations
have peak amplitudes of 10\,ppm, which matches the level inferred from
ground-based velocity observations.  We show simulations for three
different values of the mode lifetime (14, 3 and 1\,d), made using the
method described by \citet{SKB2004} and \cite{BKB2004}.

The simulations also included granulation noise at 1.5 times the solar
level (see below), which is responsible for the noise floor that rises
towards lower frequencies in the upper three panels.  In those panels, no
photon or instrumental noise was included in the simulation.  In the lower
three panels, on the other hand, we added sufficient noise to bring the
total power up to the level observed by \MOST.  As expected, we see that
shortening the mode lifetimes has the effect of lowering the peak heights,
making the oscillations even harder to detect, as also found by M04
(Supplemental Material).  It is apparent that if Procyon has oscillations
with the amplitudes suggested by velocity measurements, we would not expect
to see them clearly in the \MOST{} spectrum unless the mode lifetimes were
much longer than in the Sun.

\section{Has \MOST{} detected granulation noise?}

M04 suggested that the increase in noise towards low frequencies that is
seen in their amplitude spectrum could be caused by light variations due to
stellar granulation.  We would expect granulation to be much easier to
detect in intensity than velocity, but the level of non-white noise in the
\MOST{} data seems too high for this.

\begin{figure}
 \resizebox{\hsize}{!}{\includegraphics{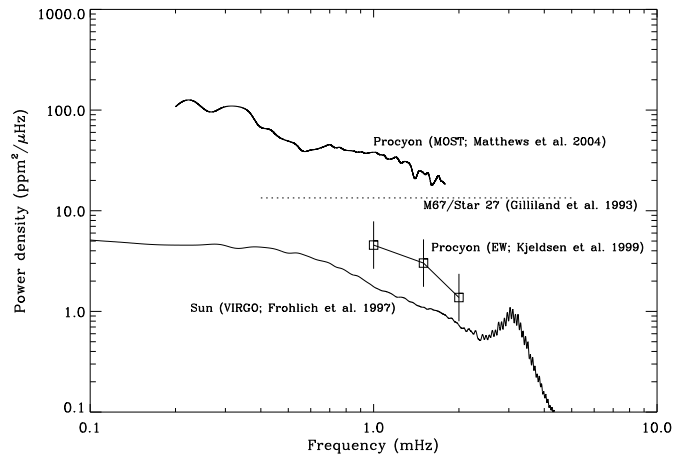}}
\caption{\label{fig.power-density} Smoothed power density spectra from
    photometry of Procyon (\MOST), the Sun (\VIRGO{}, green channel), and
    an F-type star in the cluster M67 (ground-based telescopes).  Also
    shown are Balmer-line equivalent-width measurements of Procyon (EW,
    with 2-$\sigma$ error bars).}
\end{figure}

We illustrate this in Fig.~\ref{fig.power-density}, which shows smoothed
power density spectra (power per frequency resolution bin).  Power density
is the appropriate way to compare noise sources, whether the noise arises
from granulation, photon statistics or instrumental effects, because it is
independent of the length of the observations \citep[see, e.g., Appendix
A.1 of][]{K+B95}.  The data for the Sun are full-disk intensity
measurements from the green channel of the \VIRGO{} three-channel
sunphotometer on the \SOHO{} spacecraft \citep{FAA97,PRCJ99}.  The figure
shows that the power density of the \MOST{} noise is about 16 times higher
than the solar granulation power (4 times in amplitude).  This seems
implausibly large, especially as Balmer-line equivalent width measurements
of Procyon by \citet{KBF99} produced an upper limit on granulation -- and a
tentative detection -- of about twice solar in power (1.4 in amplitude).
Those points are also shown in Fig.~\ref{fig.power-density}, after
conversion from equivalent-width to intensity (see Sec.~5 of that paper).
We also see that the noise level in the \MOST{} spectrum of Procyon is
higher than the noise level measured from photometry of F-type stars in M67
by \citet{GBK93}.%
\footnote{Note that all four power density spectra in
Fig.~\ref{fig.power-density} were scaled in the same way, such that a
sinusoid with an amplitude of 1\,ppm observed for $10^6$\,s gives a peak in
power density of height 1\,ppm$^2$/\muHz.  This explains the factor of two
between the \VIRGO{} power density in the figure and that plotted by
\citet{PRCJ99}, who used a different scaling convention.}

Given these comparisons, it seems unlikely that the noise in the \MOST{}
spectrum is due to granulation in Procyon.  We suggest that the most likely
origin of the large amount of non-white noise in the \MOST{} data is
instrumental, possibly from variations in the amount of scattered
Earthlight reaching the detector (see~M04).

\section{Could the ground-based results be artefacts?}

M04 suggested the possibility that apparent p-mode power identified by
ground-based observers may be an artefact of insufficient gapped sampling
of granulation variations.  Is it possible that the hump of excess power at
0.5--1.5\,mHz that has been seen repeatedly in the velocity data is an
artefact?  This was in fact suggested by \citet{K+B95} as a likely cause
for the power excess found by \citet{BGN91}, on the basis that the
observations were high-pass filtered.  Such a filter, when applied to a
Fourier spectrum dominated by non-white noise from instrumental effects,
will indeed produce a hump.  However, subsequent repeated independent
confirmations of the power hump, most of which were high-pass filtered much
less strongly (or not at all), led \citet{B+K2003} to acknowledge that this
excess has a stellar origin.

We also note that in discussing Fig.~7 from \citet{MSL99}, M04 have chosen
a figure showing only a subset of the data discussed in that paper, which
was deliberately cut to allow comparison with observations of another star
($\eta$~Cas).  On the other hand, Figs.~4 and~8 from \citet{MSL99} were not
high-pass filtered at all and clearly show the excess power.

The suggestion by M04 is that non-white noise from stellar granulation
could have produced a hump of power because the ground-based observations
have daily gaps.  While gaps in the observing window certainly introduce
false peaks (aliases), these are always symmetrical about the original peak
and are very close to it (small multiples of $\pm11.6$\,\muHz).  It is not
possible using this mechanism to turn a monotonically varying power
spectrum (as produced by granulation or, indeed, by most instrumental
effects) into a hump of excess power, as we have verified using
simulations.  We therefore disagree strongly that the excess power seen by
ground-based observers can be explained as an artefact introduced by gaps
in the observations.

Finally, we note that in response to M04, \citet{BMM2004} have published a
short sequence of very high-precision velocity measurements of Procyon
obtained with the HARPS spectrograph.  These show clear periodicities with
typical periods of 18 minutes and an excess of power in the Fourier
spectrum around 1\,mHz, in perfect agreement with the previous velocity
measurements.  Recent evidence for the power excess has also been presented
by \citet{CBL2005} on the basis of observations with a longer timespan but
lower precision.

\section{Evidence for oscillations in the \MOST{} data?}

We have examined the \MOST{} amplitude spectrum for evidence of
oscillations.  Based on the ground-based observations, oscillations could
be present at low signal-to-noise.  Indeed, inspection of Fig.~3 in M04
shows a few peaks that are well above the surrounding noise and it is
possible that some of these are due to oscillations.

\begin{figure}
 \resizebox{\hsize}{!}{\includegraphics{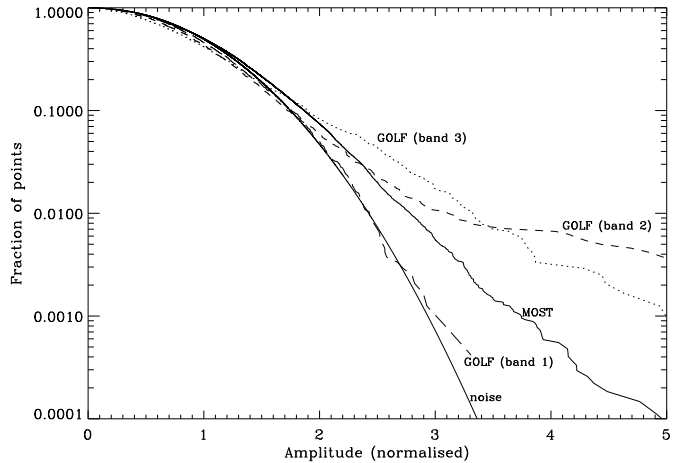}}
\caption{\label{fig.histogram} Distribution of peaks heights in the \MOST{}
    amplitude spectrum of Procyon, compared with pure noise and also with
    three frequency bands in the \GOLF{} solar spectrum (see text).}
\end{figure}

Comparison with simulations shows that there are many more strong peaks in
the amplitude spectrum than would be expected from noise.  To quantify
this, we examined the cumulative distribution of peak heights, a method
that has previously been applied in both helioseismology
\citep[e.g.,][]{D+S83} and asteroseismology \citep[e.g.,][]{BGN91}.  We
first removed the overall slope by dividing by a smoothed version of the
spectrum.  This flattens the spectrum and allows us to make meaningful
counts of the number of points above a given amplitude.  The resulting
distribution is shown in Fig.~\ref{fig.histogram}, where the horizontal
axis is amplitude divided by the mean amplitude.

On the same graph we show the distribution resulting from noise, which we
measured by averaging a large number of simulations.  It is clear that the
\MOST{} spectrum has an excess of strong peaks.  For example, there are
more than ten times as many peaks with height $\ge3.5$ times the mean than
expected from noise.

Also shown in Fig.~\ref{fig.histogram} are distributions of peak heights
for the Sun, based on a 30-day series of full-disk velocity observations
taken by the \GOLF{} instrument on the \SOHO{} spacecraft \citep{UGR2000}.
We show results for three frequency bands in the solar amplitude spectrum.
Band~1 (0.4--1.4\,mHz) is at frequencies below the solar oscillations and
contains only noise from granulation.  We see that this distribution
closely matches our noise simulations.  Band~2 (1.4--2.4\,mHz) contains
granulation noise plus the weak low-frequency solar p-modes.  Those modes
have long lifetimes compared to the length of the time series and so the
distribution contains unresolved high peaks.  Band~3 (3.8--4.8\,mHz)
contains the weak highest-frequency solar p-modes, which have short
lifetimes.  Thus, this band has many resolved modes, which produces a
broader distribution of peaks.

Comparing the different curves in Fig.~\ref{fig.histogram}, we conclude
that the distribution of peak heights in the \MOST{} amplitude spectrum is
not consistent with being solely due to noise from granulation or similar
sources.  Rather, there is evidence for an excess of strong peaks that is
consistent with the presence of oscillations at approximately the level
expected.  However, the unusual peak distribution could arise from other
causes (as turned out to be the case for the solar measurements analysed by
\citealt{D+S83}).  In particular, the filtering by M04 to remove power at
the harmonics of the satellite's orbital frequency may have distorted the
peak-height distribution and further analysis of the original time series
is needed to help decide this question.

\section{Conclusions}

Our main conclusions are as follows:
\begin{enumerate}

\item The upper limit placed on oscillations in Procyon by the \MOST{}
satellite is consistent with previous ground-based velocity measurements
and with the fact that Procyon oscillates with an amplitude significantly
below theoretical predictions, as has been discussed in the literature for
many years.  The conclusions by M04 that their results ``defy expectations
from the Sun's oscillations and previous theoretical predictions'' and that
``target selection for future planned asteroseismology space missions may
need to be reconsidered, as will the theory of stellar oscillations'' are
overstated.


\item The claims by M04 that the \MOST{} data represent the best
photometric precision so far reported in the literature by about an order
of magnitude and are the most sensitive data set for asteroseismology
available for any star other than the Sun are both inaccurate.  For
example, published observations of oscillations in \acena{} in both
photometry and velocity are far superior.

\item We find it unlikely that the non-white noise in the \MOST{} amplitude
spectrum is due to stellar granulation.  Instead, the most likely
explanation lies in variable amounts of scattered Earthlight reaching the
detector.  This instrumental noise is probably the reason that oscillations
could not be detected in Procyon.

\item The distribution of peak heights in the \MOST{} amplitude spectrum is
not consistent with a simple noise source such as granulation.  The excess
of strong peaks may indicate oscillations, but further work is needed
before drawing a firm conclusion.

\end{enumerate}

The excellent observing duty cycle of \MOST{} makes this a potentially
powerful instrument for asteroseismology.  It is possible that more
sophisticated treatment of the stray light will reduce the noise level
enough to reveal a clear detection of oscillations in Procyon.  If the
non-white noise cannot be reduced, the best targets will be those with
larger amplitudes than Procyon, such as more evolved stars (e.g.,
$\eta$~Boo) and classical pulsators such as $\delta$~Scuti stars and roAp
stars.  \MOST{} has tremendous potential to contribute greatly to the study
of these types of stars.

\begin{acknowledgements}
We thank the \MOST{} team for providing the amplitude spectrum in digital
form.  This work has been supported by the Australian Research Council and
the Danish Natural Science Research Council.
\end{acknowledgements}

\bibliographystyle{aa}

\begin{thebibliography}{40}
\expandafter\ifx\csname natexlab\endcsname\relax\def\natexlab#1{#1}\fi

\bibitem[{Barban {et~al.}(1999)Barban, {Michel}, {Marti{\'c}}, {Schmitt},
  {Lebrun}, {Baglin}, \& {Bertaux}}]{BMM99}
Barban, C., {Michel}, E., {Marti{\'c}}, M., {et~al.} 1999, A\&A, 350, 617

\bibitem[{Bedding \& Kjeldsen(2003)}]{B+K2003}
Bedding, T.~R. \& Kjeldsen, H. 2003, PASA, 20, 203

\bibitem[{Bedding {et~al.}(2004)Bedding, Kjeldsen, Butler, McCarthy, Marcy,
  O'Toole, Tinney, \& Wright}]{BKB2004}
Bedding, T.~R., Kjeldsen, H., Butler, R.~P., {et~al.} 2004, ApJ, 614, 380

\bibitem[{Bouchy \& {Carrier}(2002)}]{B+C2002}
Bouchy, F. \& {Carrier}, F. 2002, A\&A, 390, 205

\bibitem[{Bouchy \& {Carrier}(2003)}]{B+C2003}
Bouchy, F. \& {Carrier}, F. 2003, Ap\&SS, 284, 21

\bibitem[{Bouchy {et~al.}(2004)Bouchy, Maeder, Mayor, M{\'e}gevand, Pepe, \&
  Sosnowska}]{BMM2004}
Bouchy, F., Maeder, A., Mayor, M., {et~al.} 2004, Nat, 432, 7015

\bibitem[{Bouchy {et~al.}(2002)Bouchy, {Schmitt}, {Bertaux}, \&
  {Connes}}]{BSB2002}
Bouchy, F., {Schmitt}, J., {Bertaux}, J.-L., \& {Connes}, P. 2002, in IAU
  Colloqium 185: Radial and Nonradial Pulsations as Probes of Stellar Physics,
  ed. C.~Aerts, T.~R. Bedding, \& J.~Christensen-Dalsgaard, Vol. 259 (ASP Conf.
  Ser.), 472

\bibitem[{Brown {et~al.}(1991)Brown, Gilliland, Noyes, \& Ramsey}]{BGN91}
Brown, T.~M., Gilliland, R.~L., Noyes, R.~W., \& Ramsey, L.~W. 1991, ApJ, 368,
  599

\bibitem[{Butler {et~al.}(2004)Butler, Bedding, Kjeldsen, McCarthy, O'Toole,
  Tinney, Marcy, \& Wright}]{BBK2004}
Butler, R.~P., Bedding, T.~R., Kjeldsen, H., {et~al.} 2004, ApJ, 600, L75

\bibitem[{Carrier \& {Bourban}(2003)}]{C+B2003}
Carrier, F. \& {Bourban}, G. 2003, A\&A, 406, L23

\bibitem[{Chaboyer {et~al.}(1999)Chaboyer, {Demarque}, \& {Guenther}}]{CDG99}
Chaboyer, B., {Demarque}, P., \& {Guenther}, D.~B. 1999, ApJ, 525, L41

\bibitem[{Chaplin {et~al.}(1997)Chaplin, {Elsworth}, {Isaak}, {McLeod},
  {Miller}, \& {New}}]{CEI97}
Chaplin, W.~J., {Elsworth}, Y., {Isaak}, G.~R., {et~al.} 1997, MNRAS, 288, 623

\bibitem[{Christensen-Dalsgaard \& Frandsen(1983)}]{ChD+F83}
Christensen-Dalsgaard, J. \& Frandsen, S. 1983, Sol. Phys., 82, 469

\bibitem[{Claudi {et~al.}(2005)Claudi, {Bonanno}, {Leccia}, {Ventura},
  {Desidera}, {Gratton}, {Cosentino}, {Paterno}, \& {Endl}}]{CBL2005}
Claudi, R.~U., {Bonanno}, A., {Leccia}, S., {et~al.} 2005, A\&A, 429, L17

\bibitem[{Delache \& {Scherrer}(1983)}]{D+S83}
Delache, P. \& {Scherrer}, P.~H. 1983, Nat, 306, 651

\bibitem[{Di~Mauro \& {Christensen-Dalsgaard}(2001)}]{DiM+ChD2001}
Di~Mauro, M.~P. \& {Christensen-Dalsgaard}, J. 2001, in IAU Symposium 203:
  Recent Insights into the Physics of the Sun and Heliosphere: Highlights from
  SOHO and Other Space Missions, ed. P.~Brekke, B.~Fleck, \& J.~B. Gurman (ASP
  Conf. Ser.), 94

\bibitem[{Eggenberger {et~al.}(2004)Eggenberger, {Carrier}, {Bouchy}, \&
  {Blecha}}]{ECB2004}
Eggenberger, P., {Carrier}, F., {Bouchy}, F., \& {Blecha}, A. 2004, A\&A, 422,
  247

\bibitem[{Elsworth \& {Thompson}(2004)}]{E+T2004}
Elsworth, Y.~P. \& {Thompson}, M.~J. 2004, Astronomy and Geophysics, 45, 14

\bibitem[{{Fr{\"o}hlich} {et~al.}(1997){Fr{\"o}hlich}, {Andersen},
  {Appourchaux}, {et~al.}}]{FAA97}
{Fr{\"o}hlich}, C., {Andersen}, B.~N., {Appourchaux}, T., {et~al.} 1997, Sol.
  Phys., 170, 1

\bibitem[{Gilliland {et~al.}(1993)Gilliland, Brown, Kjeldsen, McCarthy, Peri,
  {et~al.}}]{GBK93}
Gilliland, R.~L., Brown, T.~M., Kjeldsen, H., {et~al.} 1993, AJ, 106, 2441

\bibitem[{Houdek {et~al.}(1999)Houdek, {Balmforth}, {Christensen-Dalsgaard}, \&
  {Gough}}]{HBChD99}
Houdek, G., {Balmforth}, N.~J., {Christensen-Dalsgaard}, J., \& {Gough}, D.~O.
  1999, A\&A, 351, 582

\bibitem[{{Kambe} {et~al.}(2003){Kambe}, {Sato}, {Takeda}, {Izumiura},
  {Masuda}, \& {Ando}}]{KST2003}
{Kambe}, E., {Sato}, B., {Takeda}, Y., {et~al.} 2003, in Asteroseismology
  Across the HR Diagram, ed. M.~J. Thompson, M.~S. Cunha, \& M.~J. P. F.~G.
  Monteiro (Kluwer), P331

\bibitem[{Kervella {et~al.}(2004)Kervella, {Th{\' e}venin}, {Morel},
  {Berthomieu}, {Bord{\' e}}, \& {Provost}}]{KTM2004}
Kervella, P., {Th{\' e}venin}, F., {Morel}, P., {et~al.} 2004, A\&A, 413, 251

\bibitem[{Kjeldsen \& Bedding(1995)}]{K+B95}
Kjeldsen, H. \& Bedding, T.~R. 1995, A\&A, 293, 87

\bibitem[{Kjeldsen {et~al.}(1999)Kjeldsen, Bedding, Frandsen, \& Dall}]{KBF99}
Kjeldsen, H., Bedding, T.~R., Frandsen, S., \& Dall, T.~H. 1999, MNRAS, 303,
  579

\bibitem[{Kurtz {et~al.}(1997)Kurtz, {van Wyk}, {Roberts}, {Marang}, {Handler},
  {Medupe}, \& {Kilkenny}}]{KvWR97}
Kurtz, D.~W., {van Wyk}, F., {Roberts}, G., {et~al.} 1997, MNRAS, 287, 69

\bibitem[{Marti{\'c} {et~al.}(2004)Marti{\'c}, {Lebrun}, {Appourchaux}, \&
  {Korzennik}}]{MLA2004}
Marti{\'c}, M., {Lebrun}, J.-C., {Appourchaux}, T., \& {Korzennik}, S.~G. 2004,
  A\&A, 418, 295

\bibitem[{Marti{\'c} {et~al.}(1999)Marti{\'c}, Schmitt, Lebrun, Barban, Connes,
  Bouchy, Michel, Baglin, Appourchaux, \& Bertaux}]{MSL99}
Marti{\'c}, M., Schmitt, J., Lebrun, J.-C., {et~al.} 1999, A\&A, 351, 993

\bibitem[{Matthews {et~al.}(2004)Matthews, Kuschnig, Guenther, Walker, Moffat,
  Rucinski, Sasselov, \& Weiss}]{MKG2004}
Matthews, J.~M., Kuschnig, R., Guenther, D.~B., {et~al.} 2004, Nat, 430, 51,
  {Erratum}: 430, 921

\bibitem[{Mosser {et~al.}(1998)Mosser, {Maillard}, {M\'ekarnia}, \&
  {Gay}}]{MMM98}
Mosser, B., {Maillard}, J.~P., {M\'ekarnia}, D., \& {Gay}, J. 1998, A\&A, 340,
  457

\bibitem[{{Pall{\' e}} {et~al.}(1999){Pall{\' e}}, {Roca Cort{\' e}s}, {Jim{\'
  e}nez}, {Golf}, \& {VIRGO Teams}}]{PRCJ99}
{Pall{\' e}}, P.~L., {Roca Cort{\' e}s}, T., {Jim{\' e}nez}, A., {Golf}, \&
  {VIRGO Teams}. 1999, in {Proc. Workshop on Stellar Structure theory and Tests
  of Convective Energy Transport}, ed. A.~Gim\'enez, E.~F. Guinan, \&
  B.~Montesinos, Vol. 173 (ASP Conf. Ser.), 297

\bibitem[{Provost {et~al.}(2002)Provost, {Marti{\' c}}, {Berthomieu}, \&
  {Morel}}]{PMB2002}
Provost, J., {Marti{\' c}}, M., {Berthomieu}, G., \& {Morel}, P. 2002, in First
  Eddington Workshop on Stellar Structure and Habitable Planet Finding, ed.
  B.~Battrick, F.~{Favata}, I.~W. {Roxburgh}, \& D.~{Galadi}, ESA SP-485, 309

\bibitem[{Retter {et~al.}(2004)Retter, {Bedding}, {Buzasi}, {Kjeldsen}, \&
  {Kiss}}]{RBB2004}
Retter, A., {Bedding}, T.~R., {Buzasi}, D.~L., {Kjeldsen}, H., \& {Kiss}, L.~L.
  2004, ApJ, 601, L95

\bibitem[{Samadi \& {Houdek}(2000)}]{S+H2000}
Samadi, R. \& {Houdek}, G. 2000, in The Third MONS Workshop: Science
  Preparation and Target Selection, ed. T.~Teixeira \& T.~R. Bedding (Aarhus:
  Aarhus Universitet), 27, available via the ADS

\bibitem[{Schou \& {Buzasi}(2001)}]{S+B2001}
Schou, J. \& {Buzasi}, D.~L. 2001, in Helio- and Asteroseismology at the Dawn
  of the Millenium, Proc. SOHO 10/GONG 2000 Workshop, ESA SP-464, 391

\bibitem[{Stein {et~al.}(2004)Stein, {Georgobiani}, {Trampedach}, {Ludwig}, \&
  {Nordlund}}]{SGT2004}
Stein, R., {Georgobiani}, D., {Trampedach}, R., {Ludwig}, H., \& {Nordlund},
  {\AA}. 2004, Sol. Phys., 220, 229

\bibitem[{Stello {et~al.}(2004)Stello, Kjeldsen, Bedding, De~Ridder, Aerts,
  Carrier, \& Frandsen}]{SKB2004}
Stello, D., Kjeldsen, H., Bedding, T.~R., {et~al.} 2004, Sol. Phys., 220, 207

\bibitem[{Toutain \& {Fr\"ohlich}(1992)}]{T+F92}
Toutain, T. \& {Fr\"ohlich}, C. 1992, A\&A, 257, 287

\bibitem[{Ulrich {et~al.}(2000)Ulrich, Garc{\'{\i}}a, Robillot, {Turck-Chi{\`
  e}ze}, {Bertello}, {Charra}, {Dzitko}, {Gabriel}, \& {Roca Cort{\'
  e}s}}]{UGR2000}
Ulrich, R.~K., Garc{\'{\i}}a, R.~A., Robillot, J.-M., {et~al.} 2000, A\&A, 364,
  799

\bibitem[{Walker {et~al.}(2003)Walker, {Matthews}, {Kuschnig}, {Johnson},
  {Rucinski}, {Pazder}, {Burley}, {Walker}, {Skaret}, {Zee}, {Grocott},
  {Carroll}, {Sinclair}, {Sturgeon}, \& {Harron}}]{WMK2003}
Walker, G., {Matthews}, J., {Kuschnig}, R., {et~al.} 2003, PASP, 115, 1023

\end{thebibliography}

\end{document}